\documentclass[letterpaper,twocolumn,10pt]{article}

\usepackage{booktabs}
\usepackage{amsmath}
\usepackage{amsthm}
\usepackage{amsfonts}

\usepackage{times}
\usepackage{graphicx}
\usepackage{url}
\usepackage{epsfig}
\usepackage{algorithm}
\usepackage{algorithmicx}
\usepackage[noend]{algpseudocode}
\usepackage{xspace}
\usepackage{multirow}
\usepackage{subcaption}
\usepackage{authblk}

\usepackage{mathptmx}  
\usepackage[kerning,spacing]{microtype} 
\usepackage{flushend} 

\usepackage{thmtools, thm-restate}
\usepackage{xcolor}

\usepackage{algorithm}
\usepackage{algpseudocode}

\newcommand{\ignore}[1]{}
\usepackage{filecontents}

\title{On Designing Machine Learning Models for Malicious Network Traffic Classification}

\author{Talha Ongun}
\author{Timothy Sakharaov}
\author{Simona Boboila}
\author{Alina Oprea}
\author{Tina Eliassi-Rad}
\affil{Northeastern University}

\newcommand{\myparagraph}[1]{\smallskip \noindent \textbf{#1}}

\begin{document}

\maketitle
\date{}

\begin{abstract}
Machine learning (ML) started to become widely deployed in cyber security settings for shortening the detection cycle of cyber attacks. To date, most ML-based systems are either proprietary or make specific choices of feature representations and machine learning models. The success of these techniques is difficult to assess as public benchmark datasets are currently unavailable.
In this paper, we provide concrete guidelines and recommendations for using supervised ML in cyber security. As a case study, we consider the problem of botnet detection from network traffic data. Among our findings we highlight that: (1) feature representations should take into consideration attack characteristics; (2) ensemble models are well-suited to handle class imbalance; (3) the granularity of ground truth plays an important role in the success of these methods.
\end{abstract} 

\section{Introduction}

A wide spectrum of threats ranging from opportunistic malicious
activities to sophisticated nation-sponsored campaigns threaten organizations from industry, academia, and government. These attacks usually result in loss of important information and affect consumers and businesses alike. Notable examples are the Equifax data breach in 2017 and the Anthem healthcare campaign in 2015 that compromised personal financial and medical records for millions of US citizens.

To date, most enterprises deploy many security controls in their environments and apply best practice (such as patching vulnerable systems, use of threat intelligence services, and endpoint scanning) to protect against cyber threats. Monitoring tools are deployed in most organizations either on the network (e.g., network intrusion-detection systems, web proxies, firewalls) or on the end hosts (e.g., anti-virus software, endpoint agents). With the availability of security logs collected by large enterprises, machine learning (ML) started to become an important defensive tool in face of increasingly sophisticated cyber attacks. ML techniques applied to network data include systems for detecting malicious domains (e.g., \cite{Notos,EXPOSURE,Antonakakis2012}), methods for detecting malware delivery (e.g.,~\cite{Nazca}) or command-and-control communication~\cite{DISCLOSURE,ExecScent,BAYWATCH,MADE}, techniques for detecting malicious web pages (e.g., \cite{ShadyPath}), and various industry products for enterprise threat detection (e.g., \cite{RSAML, EndgameML,AzureML,FireEyeML, SymantecML}).

ML has a lot of potential in shortening the malware detection cycle, but these algorithms tend to come with a number of shortcomings. In particular, Sommer and Paxson~\cite{SP10} highlighted the difficulties of using ML in operational settings for cyber security. The main limitations they identified were: (1) ML excels at supervised tasks by learning from labeled examples, while in cyber security most of the data is unlabeled. (2) ML errors (and in particular false positives) have high cost as alerts need to be investigated by security analysts. (3) Network traffic exhibits high diversity under normal operating conditions. (4) Performing sound evaluations is usually challenging due to unavailability of standard benchmark datasets.

In this paper, we describe some concrete guidelines and recommendations for using supervised ML in cyber security. As a case study, we consider the problem of botnet detection from network traffic data. We leverage a public dataset (CTU-13) which includes network traffic collected from a university campus and attacks launched on the university network. Among our findings, we highlight the following:

\begin{itemize}
\item Feature representations should take into consideration the specifics of the attacks. Among standard feature representations, we compare connection-level features (extracted directly from Bro logs) with aggregated traffic statistics and temporal features (using fixed time windows).

\item Class imbalance is a major issue that hinders the performance of simple linear models such as logistic regression.

\item Ensemble methods such as gradient boosting have built-in techniques that can handle class imbalance well. They achieve better performance at classifying malicious and benign connections compared to linear models.

\item The granularity of data labeling (ground truth) can impact the classification metrics substantially. If available, ground truth obtained at the level of individual network connections can boost the performance of supervised ML models.
\end{itemize}





\section{Background and Threat Model}

\subsection{Machine Learning for Network Traffic Classification}

Network Intrusion Detection is a highly active area of research. Traditional systems such as Snort are based on manually-generated rules for detecting well-known malware variants.

Recently, ML has proven to be valuable in augmenting rule-based systems. ML has the potential of detecting more advanced malicious activities that evade rule-based systems. Successful applications of ML to various types of network data for malware detection include:

\begin{itemize}
\item Domain reputation systems using passive DNS data, such as Notos~\cite{Notos} and EXPOSURE~\cite{EXPOSURE}.

\item Command-and-control detection based on NetFlow data, such as DISCLOSURE~\cite{DISCLOSURE} and BotFinder~\cite{BotFinder}.

\item Malicious communication detection using web-proxy logs, such as ExecScent~\cite{ExecScent}, BAYWATCH~\cite{BAYWATCH}, and MADE~\cite{MADE}.

\end{itemize}

\begin{figure}[h]
\begin{centering}
\includegraphics[width=3in]{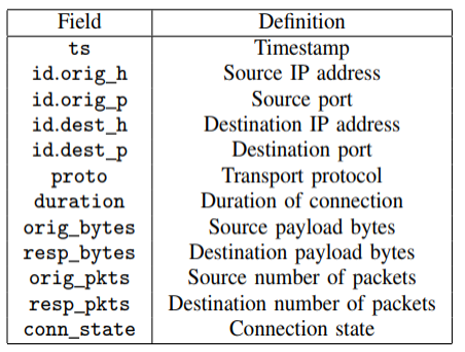}
\caption{Fields in Bro connection log.}
\label{fig:bro}
\end{centering}
\end{figure}

Bro is an open-source network monitoring agent that collects a number of network logs. Here we leverage the Bro connection logs, which record the fields included in Figure~\ref{fig:bro}. These include the TCP connection timestamp, duration, source IP and port, destination IP and port, number of packets sent and received, number of bytes sent and received, and connection state. For UDP, an entry is generated for every UDP packet (as there do not exist connections over UDP).

\subsection{Problem statement and threat model}

ML algorithms have demonstrated success in network traffic classification tasks for detecting botnets or malicious domains. However, most ML methods are designed in an ad-hoc manner and guidelines for principled approaches in this space are currently missing. We are interested in filling this gap and providing recommendations on several general principles that should guide ML design for botnet and malware detection. We are specifically addressing the problem of detecting botnets from network logs (as generated by Bro logs), but our methods can be used with other network data types (such as NetFlow, pcap, firewalls). Some of the research questions we would like to answer are the following:

\begin{itemize}
\item Can raw network data be used effectively in an ML algorithm?
\item Which feature representations are most appropriate for applying ML classification algorithms?
\item Which classifiers achieve best performance in handling the largely imbalanced cyber-security datasets?
\item What is the impact of labeling the data for ground truth generation?
\end{itemize}

We assume that the monitoring agent, which collects the network data, is not under the attacker's control. We also assume that the attacker cannot tamper with the collected network logs. Therefore, attackers do not have access to the storage device where data is recorded. \footnote{Attackers with access to the monitoring environment and the system logs are much more powerful, and are beyond our current scope.}

\section{Case Study on ML for Botnet Detection}

\begin{figure*}[t]
\begin{centering}
\includegraphics[width=5in]{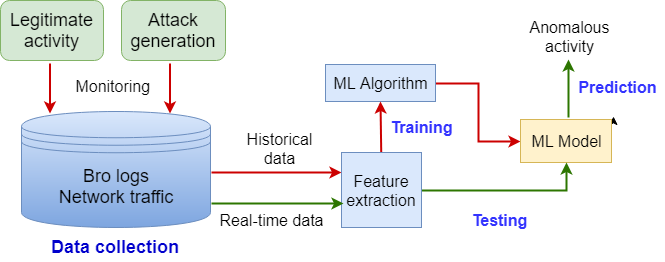}
\caption{Overview of the system architecture.}
\label{fig:system}
\end{centering}
\end{figure*}

\subsection{Dataset}

We leverage a dataset of botnet traffic that was captured in 2011 at the CTU University in the Czech Republic. The dataset includes 13 scenarios, each including legitimate traffic, as well as various attacks such as spam, port scanning, DDOS, and click fraud. The dataset also includes a list of botnet IPs that can be used for labeling the traffic.

Since ML classification needs to use similar attack data for training and testing, we decided to use a subset of 6 scenarios. Among these, 3 scenarios are generated by botnet Neris (performing spam and click fraud activity), and 3 scenarios are generated by botnet Rbot (performing DDoS activity). The statistics are in Table~\ref{tab:ctu}. For other botnets, there was only one scenario available and that precluded the use of supervised ML.

In traditional ML, cross-validation is a well-known method to evaluate the generalization of a model. $k$-fold cross-validation splits the data into $k$ partitions at random, trains a model on $k-1$ of them and evaluates it on the $k$-th partition. Splitting the logs at random produces highly-correlated data between training and testing sets. Instead, we train on two scenarios, and test on the third (independent) scenario, repeating the experiment 3 times for each of the two botnets. We have thus assurances that testing data is independent from training. This method of splitting the data into training and testing (based on independent attack scenario) is more appropriate for this setting. In other contexts, the specifics of the environment need to be taken into consideration. 


\begin{table*}
\begin{center}
\scriptsize
\begin{tabular}{|c|c|c|c|c|c|c|}
\hline

Botnet & Scenario & Attack & Botnet & Botnet & Background &  Background \\
& & & raw & aggregated & raw & aggregated \\
\hline
Neris & 1 & Spam, click fraud & 31,089 & 569 & 3,067,241 & 76,614  \\ 
& 2 & Spam, click fraud & 39,730 & 407 & 1,872,270 & 54,675 \\
& 9 & Spam, click fraud & 111,895 & 2893 & 1,689,040 &  62,970 \\
& & port scan & & \\
\hline
Rbot & 4 & ICMP, UDP & 126,438 & 122 & 869,648 & 49,041   \\
& 10 & ICMP, UDP & 10,102,210 & 741 & 988,870  & 55,160 \\
& 11 & UDP & 251,814 & 20 & 75,069 & 3169  \\
\hline
\end{tabular}
\end{center}

\caption{CTU-13 botnet scenarios.}
\label{tab:ctu}
\end{table*}

\subsection{Overview} We show our system architecture in Figure~\ref{fig:system}. Our system processes network logs collected at the border of an organization (i.e., campus or enterprise network). After data collection, a feature extraction layer is employed to prepare the data for ML training. A number of classification algorithms are used to train a classifier and optimize for standard metrics, such as precision, recall, F1 score, and AUC.  The classifiers are applied to new testing scenarios in order to evaluate their generality and predict suspicious network activity. We believe that this framework is general enough to be applicable in other environments.

\subsection{Feature extraction}

We experiment with different feature representations, as described below.

\myparagraph{Connection-level representation.} This representation extracts features directly from the raw connection logs. We consider all connections in which $\mathtt{ip}$ is either $\mathtt{id.orig\_h}$ or $\mathtt{id.dest\_h}$ and we use directly the fields from the Bro connection logs as features:
      \begin{align}
      & \mathtt{ts},\mathtt{id.orig\_h}, \mathtt{id.orig\_p},\mathtt{id.dest\_h}, \nonumber  \\
      & \mathtt{id.dest\_p},\mathtt{proto}, \mathtt{duration}, \mathtt{orig\_bytes}, \nonumber \\
      & \mathtt{resp\_bytes}, \mathtt{orig\_pkts}, \mathtt{resp\_pkts} \nonumber
      \end{align}


For categorical features (e.g., $\mathtt{proto}$) we use standard one-hot encoding. In this representation, we obtained 26 features after one-hot encoding.

\begin{table*}[h]
\begin{center}
\begin{tabular}{|c|c|c|c|}
\hline
Category & Field & Operator & Definition \\
\hline
IPs & $\mathtt{id.dest\_h}$ & Distinct &  Number of IPs communicated with per port \\
(Per port) & & Distinct &  Number of Subnets communicated with per port \\
\hline
Duration & $\mathtt{duration}$ & Sum & Total duration of connection per port  \\
(Per port) & & Min & Min duration of connection per port \\
& & Max & Max duration of connection per port \\
\hline
Bytes & $\mathtt{orig\_bytes}$ &  Sum & Total bytes sent by $\mathtt{ip}$ per port \\
(Per port) & & Min &  Min bytes sent by $\mathtt{ip}$ in a connection per port \\
& & Max &  Max bytes sent by $\mathtt{ip}$ in a connection per port\\
& $\mathtt{resp\_bytes}$ & Sum & Total bytes received by $\mathtt{ip}$ per port \\
& & Min &  Min bytes received by $\mathtt{ip}$ in a connection per port\\
& & Max &  Max bytes received by $\mathtt{ip}$ in a connection per port\\
\hline
Packets & $\mathtt{orig\_pkts}$ &  Sum & Total packets sent by $\mathtt{ip}$ per port \\
(Per port) & & Min &  Min packets sent by $\mathtt{ip}$ in a connection per port\\
& & Max &  Max packets sent by $\mathtt{ip}$ in a connection per port\\
& $\mathtt{resp\_pkts}$ & Sum & Total packets received by $\mathtt{ip}$  per port\\
& & Min &  Min packets received by $\mathtt{ip}$ in a connection per port \\
& & Max &  Max packets received by $\mathtt{ip}$ in a connection per port \\
\hline
Traffic statistics & $\mathtt{proto}$ & Sum & Number of connections per transport protocol (TCP, UDP, ICMP) \\
& $\mathtt{id.orig\_p}$ & Distinct & Number of source ports \\
& $\mathtt{id.dest\_h}$ & Distinct & Number of external destination IPs \\
& $\mathtt{id.dest\_p}$ & Distinct & Number of destination ports \\
\hline
\end{tabular}
\end{center}
\caption{Traffic features aggregated by time. The top 4 categories of features are defined per port. }
\label{tab:features_traffic}
\end{table*}

\myparagraph{Aggregated traffic statistics.} Next, we would like to explore if features obtained by time aggregation are more powerful than raw features. We consider a time interval of length $T$ over which we define aggregated features over all connections in which $\mathtt{ip}$ is either $\mathtt{id.orig\_h}$ or $\mathtt{id.dest\_h}$.

An important consideration when defining our features is to generate a fixed number of features, independent of the traffic at a particular host. In our first attempt, we consider the set of all destination IP addresses that $\mathtt{ip}$ communicates with: $S_{IP} = \{IP_1,\dots,IP_n\}$. From these we can define the set of /24 destination subnets that $\mathtt{ip}$ communicates with: $S_{subnet} = \{Sub_1,\dots,Sub_m\}$, with $m \le n$. If we define aggregated features per destination or subnet, we will encounter an issue when a host visits new IPs or new destinations. In that case, we need to add new features to our representation, which is not desirable in practice.

To alleviate this problem, we  define our aggregated features by destination port (corresponding to applications or network services). Specifically, we define a set of 17 popular application ports (e.g., HTTP - 80, HTTPS - 443, SSH - 22, DNS - 53). We then take a modular approach. We select a small number of operators (Distinct, Sum, Min, Max) and apply them to fields in {\sf conn.log} for each destination port. The features are described in Table~\ref{tab:features_traffic}. We generate these features separately for outgoing and incoming connections. Additionally, we add some features that capture communication patterns with external IP destinations (e.g., number of connections per transport protocol, number of source and destination ports, number of destination IPs, etc.). In this representation, we obtain 756 aggregated traffic features.


\myparagraph{Temporal features.} Considering the same time interval $T$ as with the aggregated connection-level features, we define inter-arrival features on a node as the mean, standard deviation, median, minimum, and maximum of the time distribution between node communications.  Each internal node has two such sets of features: one for events where the node serves as the source of communication ({\bf outgoing}), and one where it is the target ({\bf incoming}).  These communications are aggregated by common ports.  Thus, in each time interval $T$, a node $i$ will have the  inter-arrival features listed in Table~\ref{tab:features_time}. In this representation, we obtain 180 features.

\ignore{
\begin{align}
& \mathtt{mean\_send\_iet}[i] \text{ for each port} \nonumber  \\
      & \mathtt{std\_send\_iet}[i] \text{ for port p1, p2, ...} \nonumber  \\
      & \mathtt{median\_send\_iet}[i] \text{ for port p1, p2, ...} \nonumber  \\
      & \mathtt{min\_send\_iet}[i] \text{ for port p1, p2, ...} \nonumber  \\
      & \mathtt{max\_send\_iet}[i] \text{ for port p1, p2, ...} \nonumber  \\
      & \mathtt{mean\_receive\_iet}[i] \text{ for port p1, p2, ...} \nonumber  \\
      & \mathtt{std\_receive\_iet}[i] \text{ for port p1, p2, ...} \nonumber  \\
      & \mathtt{median\_receive\_iet}[i] \text{ for port p1, p2, ...} \nonumber  \\
      & \mathtt{min\_receive\_iet}[i] \text{ for port p1, p2, ...} \nonumber  \\
      & \mathtt{max\_receive\_iet}[i] \text{ for port p1, p2, ...} \nonumber  \\
      \end{align}
 }

\begin{table*}[h]
\begin{center}
\begin{tabular}{|c|c|c|c|}
\hline
Category & Statistics &  Definition \\
\hline
Outgoing & Mean, std. dev., median, min, max & Statistics of inter-arrival distribution for outgoing traffic \\
Incoming & Mean, std. dev., median, min, max & Statistics of inter-arrival distribution for incoming traffic  \\
\hline
\end{tabular}
\end{center}
\caption{Temporal features aggregated by time. Each of these features is defined per port.}
\label{tab:features_time}
\end{table*}


\subsection{ML classification and labeling}

\myparagraph{Ground truth labeling} CTU-13 dataset provides a list of botnet IP addresses. One of our main observations is that the attack is not active during the duration of the entire data collection. We found that the granularity at which we label the data plays a large role in the results. We experiment with two levels of granularity:

\begin{itemize}
\item {\bf Coarse-grained labeling}: We label all the connection logs generated by the botnet IPs as {\bf Malicious} during the entire scenario period.

\item {\bf Fine-grained labeling}: For the Rbot attack (an instance of DDoS), we obtain the IP address of the victim machine. We use that to identify the attack flows that connect to the victim IP. For all feature representations, we label a time window as {\bf Malicious} if there is at least one attack log event in that time window.

\end{itemize}

Fine-grained labeling is difficult to obtain in general because it is a manual process, but when it is available it improves significantly the performance of ML in botnet detection.

\myparagraph{ML models} We consider several well-known ML classification models, including logistic regression, random forest, and gradient boosting. We use several metrics to evaluate the performance of the ML algorithms (precision, recall, F1 score, and AUC). As the imbalance is quite large in this dataset (the ratio of {\bf Malicious} to {\bf Legitimate} samples is as low as 1:134 for Neris and 1:401 for Rbot with features aggregated at 30-second intervals), the accuracy is always quite high (above 0.96 in all our experiments). We are interested in results on the minority ({\bf Malicious}) class, thus precision, recall, F1 score, and AUC are better indicators of how the classifiers perform at detecting botnets.

For the ML classifiers, we perform a grid search on several hyper-parameters to select the models performing best in our setting. For Random Forest, we selected the number of trees in $\{10,50,100,200\}$ and found that 100 tree worked best. For Gradient Boosting, we varied the number of estimators in $\{50,100,200\}$, the maximum depth of each tree in $\{3,5,7\}$ and learning rate in $\{0.01,0.05,0.1\}$. We selected 100 estimators with maximum depth of 3 and learning rate 0.05. For logistic regression, we used $L_1$ or Lasso regularization to reduce the space dimension.

\section{Experimental Evaluation}


During our experimental evaluation, we would like to answer several research questions, which we detail below.

\myparagraph{Which feature representation performs best?} We compare different feature representations (connection-level representation, aggregated traffic statistics, and temporal features). For this experiment, we use a random forest classifier with 100 trees and a 30-second time window for aggregation.

The results for Neris are in Table~\ref{tab:neris_features} and they show that aggregated features (both traffic statistics and temporal) perform significantly better than raw features extracted directly from Bro logs at all metrics of interest. For instance, when training on scenarios 2 and 9 and testing on scenario 1, the F1 score for connection features is 0.65, while the F1 score for aggregated features is 0.98. We do not observe a major difference when we consider both traffic and timing features, compared to using only aggregated traffic features.

The results for Rbot for fine-grained labeling are in Table~\ref{tab:ddos_features}. Here, connection-based features perform quite well. The reason is that this is a DDoS attack in which all packets sent to the victim are identical. However, traffic statistics and temporal features also perform well. The exception is when training on scenarios 4 and 11, and testing on scenario 10. In that case, the amount of botnet samples used for training with 30-second aggregation is very small (142), while there are much more botnet samples in the raw data (378,252).


\begin{table}
\begin{center}
\scriptsize
\begin{tabular}{|c|c|c|c|c|c|c|}
\hline

Features &  Training & Testing & Prec. & Recall & F1 & AUC  \\
& Scenarios & Scenario & & & & \\
\hline
Connection & 2,9 & 1 &  0.68 &	0.62 &	0.65 &	0.87	 \\
& 1,9 & 2 &  0.89 & 0.43	& 0.58	& 0.88 \\
& 1,2 & 9 &  0.92 &	0.70 &	0.80 &	0.94 \\
\hline
Traffic & 2,9 & 1 & {\bf 0.99} &	0.98 &	{\bf 0.98} &	{\bf 0.99} \\
& 1,9 & 2 & 0.94 &	{\bf 0.96} & {\bf 0.95} &	{\bf 0.99} \\
& 1,2 & 9 & {\bf 1} &	{\bf 0.90}	& {\bf 0.94} &	{\bf 0.96} \\
\hline
Traffic and & 2,9 & 1 & {\bf 0.99}	& 0.97	& {\bf 0.98}	& {\bf 0.99}\\
Temporal & 1,9 & 2 & {\bf  0.95} &	{\bf 0.96} &	{\bf 0.95} &	0.98 \\
& 1,2 & 9 & {\bf 1} &	{\bf 0.90} &	{\bf 0.94} &	{\bf 0.96} \\

\hline
\end{tabular}
\end{center}

\caption{Classification metrics for the Neris botnet for Random Forest with different feature representations. Best results are highlighted in bold.}
\label{tab:neris_features}
\end{table}

\begin{table}
\begin{center}
\scriptsize
\begin{tabular}{|c|c|c|c|c|c|c|}
\hline

Features &  Training & Testing & Prec. & Recall & F1 & AUC  \\
& Scenarios & Scenario & & & & \\
\hline
Connection & 10,11 & 4 & 0.99 &	0.99 &	0.99 &	0.99 \\
& 4,11 & 10 &  0.99 &	{\bf 0.99} &	{\bf 0.99} & {\bf 0.99} \\
& 4,10 & 11 & 0.99 & 0.99 &	0.99 &	0.99  \\
\hline
Traffic & 10,11 & 4 & {\bf 1} & {\bf 1} &{\bf 1} & {\bf 1} \\
& 4,11 & 10 & 1	& 0.85	& 0.92	& 0.92 \\
& 4,10 & 11 & {\bf 1} & {\bf 1} &{\bf 1} & {\bf 1} \\
\hline
Traffic and & 10,11 & 4 & {\bf 1} & {\bf 1} & {\bf 1} & {\bf 1} \\
Temporal & 4,11 & 10 &  {\bf 1}	& 0.85	& 0.92	& 0.92  \\
& 4,10 & 11 & {\bf 1} & {\bf 1} & {\bf 1} & {\bf 1} \\
\hline
\end{tabular}
\end{center}

\caption{Classification metrics for the Rbot botnet for Random Forest with different feature representations and fine-grained labeling. Best results are highlighted in bold.}
\label{tab:ddos_features}
\end{table}





\begin{figure}[th]
\centering
\includegraphics[width=.9\columnwidth]{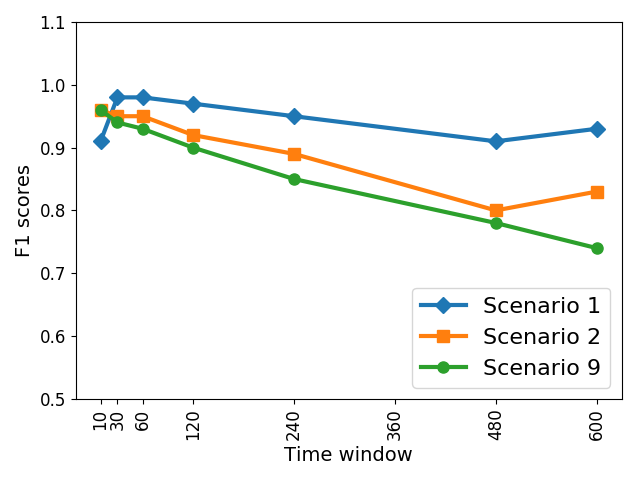}
\caption{F1 scores for different time windows for Neris.}
\label{fig:window}
\end{figure}




\begin{table}[bth]
\begin{center}
\scriptsize
\begin{tabular}{|c|c|c|c|c|c|c|}
\hline

Time &  Training & Testing & Prec. & Recall & F1 & AUC  \\
(seconds) & Scenarios & Scenario & & & & \\
\hline
1 & 2,9 & 1 & 0.89 &	0.87 &	0.88 &	0.98\\
& 1,9 & 2 & 0.92 &	0.87 &	0.90 &	0.98\\
& 1,2 & 9 & 0.98 &	0.89 &	0.93 &	{\bf 0.98} \\
\hline
10 & 2,9 & 1 & 0.84	& 0.98	& 0.91	& {\bf 0.99} \\
& 1,9 & 2 & {\bf 0.96} & {\bf 0.96} &	{\bf 0.96} & {\bf 0.99} \\
& 1,2 & 9 & {\bf 1} & {\bf 0.92}	& {\bf 0.96}	& {\bf 0.98} \\
\hline
30 & 2,9 & 1 & {\bf 0.99} &	{\bf 0.97} & {\bf 0.98} & {\bf 0.99} \\
& 1,9 & 2 & 0.95 &	{\bf 0.96} &	0.95 & 0.98 \\
& 1,2 & 9 & {\bf 1} & 0.90 &	0.94 & 0.96 \\
\hline
60 & 2,9 & 1 & {\bf 0.99} &	{\bf 0.97} &	{\bf 0.98} & {\bf 0.99} \\
& 1,9 & 2 & 0.95 &	{\bf 0.96} &	0.95 & 0.98  \\
& 1,2 & 9 & {\bf 1} &	0.87	& 0.93	& 0.95 \\
\hline
120 & 2,9 & 1 & 0.97 &	{\bf 0.97} &	0.97 &	{\bf 0.99} \\
 & 1,9 & 2 &  0.91 &	0.94	& 0.92	& 0.98 \\
& 1,2 & 9 & 0.99 &	0.82 &	0.90 &	0.92 \\
\hline
240 & 2,9 & 1 &  0.94 &	0.97 &	0.95 &	{\bf 0.99} \\
 & 1,9 & 2 & 0.85 &	0.92 &	0.89 &	0.97   \\
& 1,2 & 9 & {\bf 1} &	0.75 &	0.85 &	0.89 \\
\hline
600 & 2,9 & 1 & 0.87 &	{\bf 1} &	0.93 &	{\bf 0.99} \\
 & 1,9 & 2 & 0.76 &	0.90 &	0.83 &	{\bf 0.99} \\
& 1,2 & 9 & {\bf 1} &	0.59 &	0.74 &	0.82 \\

\hline
\end{tabular}
\end{center}
\caption{Classification metrics for the Neris botnet for Random Forest with different time windows. We used the aggregated traffic statistics and temporal features. Best results are highlighted in bold.}
\vspace{-0.5cm}
\label{tab:neris_time}
\end{table}

\myparagraph{What is the impact of varying the time window?} Here, we validate the choice of the time window for aggregation. Table~\ref{tab:neris_time} and Figure~\ref{fig:window} show results for varying the time window from 1 to 600 seconds. The 30-second and 60-second time windows exhibit similar results and they are performing well most of the time. Window size 10 is also performing well, except when testing on scenario 1.  As the time window increases beyond 120 seconds, the results start to degrade. We suspect this is because of the small samples of attack traffic at larger aggregation windows, as well as additional noise in the legitimate traffic. In general, selecting the best time window for aggregation is attack-dependent. We recommend the use of cross-validation for selecting the optimal value of the time window. Based on these results, we select a time window of 30 seconds for the rest of experiments.

\myparagraph{What is the impact of different ML models?} One important observation is that the amount of imbalance in cyber security is very large (as also observed by previous work~\cite{Bartos16,MADE}). It is well-known that ensemble classifiers such as random forests and boosting handle imbalance much better than simpler models. We test this hypotheses by using three different classifiers for our task: logistic regression, random forests, and gradient boosting. We fix the aggregation time window to 30 seconds and use the traffic statistics and temporal features.

\begin{figure}[t]
\centering
\includegraphics[width=.9\columnwidth]{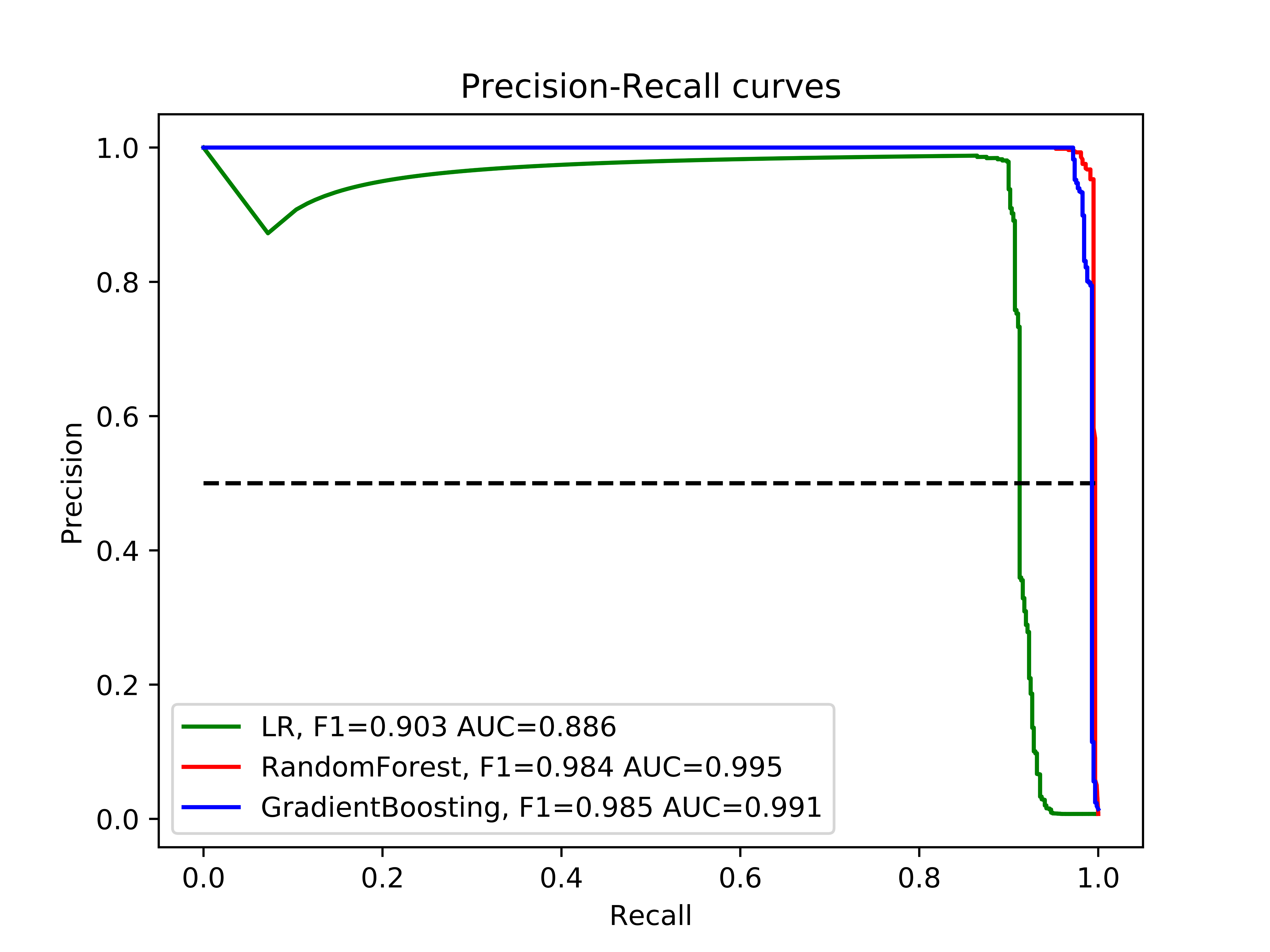}
\caption{Precision-recall curves for three classifiers for Neris.}
\label{fig:precneris}
\end{figure}

The results for three classifiers for Neris are in Table~\ref{tab:neris_models} and the precision-recall curves are in Figure~\ref{fig:precneris}. All three models we experimented with perform relatively well. Both ensemble method perform  better than the logistic regression model, with F1 scores reaching between 0.94 and 0.98 on all scenarios. The difference between random forest and gradient boosting is imperceptible, they are both powerful classification models.


\begin{table}[tbh]
\begin{center}
\scriptsize
\begin{tabular}{|c|c|c|c|c|c|c|}
\hline

Model &  Training & Testing & Prec. & Recall & F1 & AUC  \\
& Scenarios & Scenario & & & & \\
\hline
Logistic & 2,9 & 1 & 0.90 &	0.90 &	0.90 &	0.94 \\
Regression& 1,9 & 2 & 0.98 &	0.95 &	0.97 &	{\bf 0.99} \\
& 1,2 & 9 & 0.97 &	0.87 &	0.92 &	0.96 \\
\hline
Random & 2,9 & 1 & {\bf 0.99} & {\bf 0.97} &	{\bf 0.98} &	{\bf 0.99}\\
Forest & 1,9 & 2 & 0.95 &	{\bf 0.96} &	0.95 &	0.98 \\
& 1,2 & 9 & {\bf 1} &	{\bf 0.90} &	{\bf 0.94} &	{\bf 0.96} \\
\hline
Gradient & 2,9 & 1 & {\bf 1} &	{\bf 0.97} & {\bf 0.98} &	{\bf 0.99} \\
boosting & 1,9 & 2 & {\bf 1} &	0.92 &	{\bf 0.96} &	{\bf 0.99} \\
& 1,2 & 9 & {\bf 1} &	0.87 &	0.93 &	0.95 \\

\hline
\end{tabular}
\end{center}

\caption{Classification metrics for the Neris botnet for three classifiers for aggregated traffic statistics and temporal features (aggregation window 30 seconds). Best results are highlighted in bold.}
\label{tab:neris_models}
\end{table}

\myparagraph{Are the models interpretable?} To understand what the ML models learned, we computed feature importance for the random forest classifier for both Neris and Rbot (using the aggregated traffic statistics and timing features at 30-second window). The results are in Table~\ref{tab:fimp}. Interestingly, we observe that the classifier identifies features that are correlated with the attack. Neris is a spam botnet and most of its activity uses port 25, making features such as distinct source ports and median inter-arrival packet time on port 25 most relevant. In contrast, Rbot is a DDoS botnet that uses different ports for the attack. For instance, the UDP flood is using port 161, and the classifier correctly determines that the standard deviation of inter-arrival packet timing on port 161 is the most important feature.

These results show our framework's flexibility and ability to generalize to different attack patterns. We defined a set of 936 generic features that can be used for a variety of botnet attacks. For the two different botnets we experimented with, the ML models identified the most relevant features that are correlated with the attacks, without the need for a human expert to explicitly locate those features. Models such as random forest provide standard metrics for feature importance, with a clear advantage for model interpretability compared to deep learning and neural networks that lack interpretability. Interpretability is important in cyber security, as most of the time human experts analyze the alerts of ML systems. 

\begin{table}
\begin{center}
\scriptsize
\begin{tabular}{|c|c|c|c|}
\hline
Botnet & Feature & Port & Importance \\
\hline
Neris & Distinct source ports & 25  & 0.085 \\
& Median inter-arrival time  & 25 & 0.070 \\
& Distinct destination ports &  25  & 0.067 \\	
& Min packets sent & 25 &   0.061	\\
& Distinct external IPs & 25 &  0.054	\\
& Total duration & 25  & 0.053	\\
& Total packets sent & 25  & 0.051	\\
& Max duration of connection & 25  & 0.048	\\
\hline
\hline
Rbot & Std. dev. of inter-arrival time & 161 & 0.049 \\
& Distinct source ports & 135 & 0.046	\\
& Distinct source ports & Other & 0.043 \\	
& Min inter-arrival timing & 138 &  0.042	\\
& Distinct source ports & 138 &  0.040	\\
& Distinct source ports & 3 &  0.038	\\
& Distinct source ports & 8 &  0.030	\\
& Std. dev. of inter-arrival time & 138 & 0.029 \\	
\hline
\end{tabular}
\end{center}

\caption{Feature importance for the Neris botnet (top) and the Rbot botnet (bottom). All these features are for outgoing connections from an internal node.}
\vspace{-0.5cm}
\label{tab:fimp}
\end{table}

\myparagraph{What is the impact of labeling flows accurately?}
We perform an experiment to test how the granularity of data labeling impacts the classification results. For the Rbot DDoS botnet we have access to the IP address of the victim machine and thus we can determine which connections are botnet-related. We refer to \emph{fine-grained} labeling to  the process of labeling only the botnet connection to victim IP as {\bf Malicious}. We refer to coarse-grained labeling to the process of labeling all connections initiated by the botnet IP as {\bf Malicious}.

Table~\ref{tab:rbot_label} shows the results of fine-grained and coarse-grained labeling for the Random Forest and Gradient Boosting classifiers for features aggregated at 30-second intervals. The results demonstrate that classifier performance obtained with fine-grained labeling is much better than using coarse-grained labeling. For instance, when training on scenarios 10 and 11, and testing on scenario 4, the F1 score for coarse-grained labeling is 0.44, compared to a perfect F1 score for fine-grained labeling.  Both classifiers perform here similarly for fine-grained labeling.

\begin{table}
\begin{center}
\scriptsize
\begin{tabular}{|c|c|c|c|c|c|c|c|}
\hline

Model & Label & Training & Testing & Prec. & Recall & F1 & AUC  \\
& & Scenarios & Scenario & & & &  \\
\hline
Random & C & 10,11 & 4 & 0.75 &	0.31 &	0.44 &	0.94 \\
Forest & & 4,11 & 10 & 0.99 &	0.76 &	0.86 &	0.90 \\
& & 4,10 & 11 & 0.93 &	0.75 &	0.83 &	0.91 \\

\hline
Random & F & 10,11 & 4 & {\bf 1} &	{\bf 1} &	{\bf 1} & {\bf 1} \\
Forest & & 4,11 & 10 &  {\bf 1}	& {\bf 0.85}	& {\bf 0.92} & 0.92  \\
& & 4,10  & 11 & {\bf 1} &	{\bf 1} &	{\bf 1} &	{\bf 1}  \\

\hline
Gradient & {\bf C} & 10,11 & 4 & {\bf 1} &	0.27 &	0.42 &	0.84 \\
boosting & & 4,11 & 10 &  0.99 &	0.74 &	0.84 &	{\bf 0.95} \\
& & 4,10 & 11 & {\bf 1} &	0.75 &	0.85 &	0.92 \\

\hline
Gradient & {\bf F} & 10,11 & 4 & {\bf 1} &	{\bf 1} &	{\bf 1} &	{\bf 1} \\
boosting & & 4,11 & 10 & {\bf  1} &	{\bf 0.85}	& {\bf 0.92} & 0.92 \\
& & 4,10 & 11 & {\bf 1} &	{\bf 1} &	{\bf 1} &	{\bf 1} \\

\hline
\end{tabular}
\end{center}

\caption{Classification metrics for the Rbot botnet for two methods of labeling the data (coarse-grained or {\bf C} and fine-grained or {\bf F}). We used the Random Forest (100 trees) and Gradient Boosting classifiers for aggregated traffic statistics and temporal features (aggregation window 30 seconds). Best results are highlighted in bold.}
\label{tab:rbot_label}
\end{table}


\section{Lessons and General Recommendations}


Motivated by our case study of botnet classification from Bro logs, we highlight several guidelines that we believe are applicable in other settings where ML is used in cyber security.

\myparagraph{Multiple feature representations need to be evaluated.} Features extracted directly from raw data such as Bro connection logs do not always results in the most optimal representation. A representation that worked well in our setting for classifying internal IP addresses is  feature aggregation by time windows and port number. We also observed that feature representation depends on the amount of training data available. With the large imbalance between the malicious and benign classes, smaller time windows work better for aggregation. However, the right feature representation and the choice of time window for feature aggregation are dependent on the attack type. We recommend evaluating multiple feature representations.

\myparagraph{Model interpretability.} Models that provide interpretability are preferred in cyber security as security analysts need to investigate the alerts raised by ML systems. Understanding why a flow is labeled as malicious can speed up the investigation significantly. We showed how a random forest classifier is interpretable by identifying most relevant features that clearly provide insights about the botnet activity.

\myparagraph{Data imbalance raises a challenge for supervised learning.} Data imbalance results in a huge challenge when applying classification methods to cyber security. Simpler models such as linear models are not equipped to deal well with class imbalance. We showed that ensemble models such as random forest and gradient boosting achieve good results even in highly imbalanced scenario, compared to logistic regression. For instance, at an imbalance of 1:134 (when testing on scenario 2 for Neris) we obtain 0.97 precision and 0.95 recall with gradient boosting.

The alternative to classification is to employ anomaly-detection models that learn from the legitimate class and identify attacks as anomalies. Nevertheless, Sommer and Paxson~\cite{SP10} discussed extensively the difficulty of using anomaly detection in cyber security. We plan to investigate the performance of anomaly detectors in future work.

\myparagraph{Fine-grained ground truth labeling can be a major factor in the success of supervised learning.}
As we demonstrated, data labeling for generating the ground truth plays a major factor in measuring the success of supervised learning algorithms. If detailed information about the attack is available (e.g., the destination IPs contacted by attacker), then the performance of classifiers can be greatly improved. However, it is difficult most of the time to identify exactly the attack flows, even when running cotrolled attack simulations. Malware can contact a variety of IP addresses using different protocols, but infected machines also generate a fair number of legitimate connections (e.g., connections to Window updates).




\section*{Acknowledgements}

The research reported in this document/presentation was performed in connection with contract number W911NF-18-C-0019 with the U.S. Army Contracting Command - Aberdeen Proving Ground (ACC-APG) and the Defense Advanced Research Projects Agency (DARPA). The views and conclusions contained in this document/presentation are those of the authors and should not be interpreted as presenting the official policies or position, either expressed or implied, of ACC-APG, DARPA, or the U.S. Government unless so designated by other authorized documents. Citation of manufacturer's or trade names does not constitute an official endorsement or approval of the use thereof. The U.S. Government is authorized to reproduce and distribute reprints for Government purposes notwithstanding any copyright notation hereon. 

   We thank Malathi Veeraraghavan, Jack Davidson, Alastair Nottingham, and Donald Brown from University of Virginia, Kolia Sadeghi from Commonwealth Computer Research, Inc., and other PCORE-project team members for their support of this work. We would also like to thank Vijay Sarvepalli, Andrew J Kompanek, and Lena Pons from the Software Engineering Institute at Carnegie Mellon University for their helpful feedback regarding the evaluation.

\bibliographystyle{abbrv}
\bibliography{refs}

\end{document}